\documentclass[twocolumn,aps,superscriptaddress,showpacs,floatfix]{revtex4-1}
\usepackage{graphicx}
\usepackage{dcolumn}
\usepackage{bm}
\usepackage{booktabs}
\usepackage{color}
\usepackage{cancel}
\usepackage{array}
\usepackage{soul}
\usepackage{multirow}
\usepackage{amsmath,amssymb}  
\usepackage{threeparttable}  
\usepackage{graphicx}
\usepackage{lineno}
\usepackage{ulem} 
\usepackage{xcolor} 

\usepackage[colorlinks,linkcolor=blue,anchorcolor=blue,citecolor=blue,urlcolor=blue]{hyperref}

\begin{document}

\title{Role of nuclear and electromagnetic fragmentation in the charge-changing reactions of $^{18}$O on carbon and lead targets at around 370 MeV/nucleon}
\author{J.R. Liu}
\affiliation{School of Physics, Beihang University, Beijing, 100191, China}

\author{B.-H. Sun}
\email{Corresponding author: bhsun@buaa.edu.cn}
\affiliation{School of Physics, Beihang University, Beijing, 100191, China}

\author{J.W. Zhao}
\email{Corresponding author: zhaojianwei@buaa.edu.cn}
\affiliation{School of Physics, Beihang University, Beijing, 100191, China}

\author{G. Guo}
\affiliation{School of Physics, Beihang University, Beijing, 100191, China}

\author{G.S. Li}
\affiliation{School of Physics, Beihang University, Beijing, 100191, China}

\author{Z.Z. Li}
\affiliation{School of Physics, Peking University, Beijing, 100871, China}

\author{Y.F. Niu}
\affiliation{School of Nuclear Science and Technology, Lanzhou University, Lanzhou 730000, China}
\affiliation{Frontiers Science Center for Rare isotope, Lanzhou University, Lanzhou 730000, China}

\author{I. Tanihata} 
\affiliation{School of Physics, Beihang University, Beijing, 100191, China}
\affiliation{RCNP, Osaka University, Ibaraki Osaka 567-0047, Japan}

\author{S. Terashima}
\affiliation{School of Physics, Beihang University, Beijing, 100191, China}
\affiliation{Institute of Modern Physics, Chinese Academy of Sciences, Lanzhou, 730000, China}

\author{F. Wang}
\affiliation{School of Physics, Beihang University, Beijing, 100191, China}

\author{M. Wang}
\affiliation{School of Physics, Beihang University, Beijing, 100191, China}

\author{X.L. Wei}
\affiliation{School of Physics, Beihang University, Beijing, 100191, China}

\author{J.Y. Xu}
\affiliation{School of Physics, Beihang University, Beijing, 100191, China}

\author{J.C. Zhang}
\affiliation{School of Physics, Beihang University, Beijing, 100191, China}

\author{L.H. Zhu}
\affiliation{School of Physics, Beihang University, Beijing, 100191, China}

\author{L.C. He}
\affiliation{School of Physics, Beihang University, Beijing, 100191, China}

\author{C.Y. Liu}
\affiliation{School of Physics, Beihang University, Beijing, 100191, China}

\author{C.G. Lu}
\affiliation{Institute of Modern Physics, Chinese Academy of Sciences, Lanzhou, 730000, China}

\author{W.J. Lin}
\affiliation{School of Physics, Beihang University, Beijing, 100191, China}

\author{W.P. Lin}
\affiliation{Key Laboratory of Radiation Physics and Technology of the Ministry of Education, Institute of Nuclear Science and Technology, Sichuan University, Chengdu, 610064, China}

\author{Z. Liu}
\affiliation{Institute of Modern Physics, Chinese Academy of Sciences, Lanzhou, 730000, China}
\affiliation{School of Nuclear Science and Technology, University of Chinese Academy of Sciences, Beijing, 100049, China}

\author{P.P. Ren}
\affiliation{Key Laboratory of Radiation Physics and Technology of the Ministry of Education, Institute of Nuclear Science and Technology, Sichuan University, Chengdu, 610064, China}

\author{Y.Z. Sun}
\affiliation{Institute of Modern Physics, Chinese Academy of Sciences, Lanzhou, 730000, China}

\author{Z.Y. Sun}
\affiliation{Institute of Modern Physics, Chinese Academy of Sciences, Lanzhou, 730000, China}

\author{J. Wang}
\affiliation{School of Physics, Beihang University, Beijing, 100191, China}

\author{S.T. Wang}
\affiliation{Institute of Modern Physics, Chinese Academy of Sciences, Lanzhou, 730000, China}

\author{X.D. Xu}
\affiliation{Institute of Modern Physics, Chinese Academy of Sciences, Lanzhou, 730000, China}

\author{M.X. Zhang}
\affiliation{School of Physics, Beihang University, Beijing, 100191, China}

\author{X.H. Zhang}
\affiliation{Institute of Modern Physics, Chinese Academy of Sciences, Lanzhou, 730000, China}

\author{Y. Zhang}
\affiliation{School of Physics, Beihang University, Beijing, 100191, China}

\date{\today}

\begin{abstract}
Charge-changing cross sections (CCCSs) of $^{18}$O on carbon (C) and lead (Pb) targets have been measured with an uncertainty of less than 4\% at around 370~MeV/nucleon. We evaluate the contributions of nucleon-nucleon (NN) and electromagnetic (EM) interactions to CCCSs by considering the direct proton removal process, the charged particle evaporation (CPE) after neutron removal, and the EM excitation.  
 We conclude that the CPE accounts for 12.3\% and 5\% of CCCSs on C and Pb, respectively. Only less than 1\% of CCCSs of $^{18}$O is attributed to the EM excitation. Further investigation of projectiles from $^{18}$O to $^{197}$Au on C, silver (Ag) and Pb targets at 300 and 900~MeV/nucleon show that the contribution of EM to CCCSs on Ag and Pb increases with projectile mass numbers and incident energies, and can reach 10\% for $^{197}$Au on Pb at 900~MeV/nucleon. In contrast, the EM contribution to CCCS is negligible for all projectiles on C at both energies.
\end{abstract}

\maketitle

\section{Introduction}
\label{intro}

The charge-changing cross section (CCCS) represents the probability of removing one or more protons from an incident nucleus upon interaction with target nuclei. Systematic CCCS measurements have enabled the development of empirical formulas/models to predict these cross sections and contribute to applications in medical physics and space science~(see, e.g., Refs.~\cite{Abdulmagead2025,LiGS2023,Chulkov2000,Ferrando1988,Horst2017,Luoni2021}). 
Furthermore, precise CCCS measurements serve as an efficient approach to look for the first hint of structural changes~\cite{Bochkarev1998,Blank1992,DingMQ2024} such as neutron skin and neutron halo, to explore the equation of state~\cite{XuJY2022} and the reaction mechanisms involved in single-nucleon removal~\cite{LiGS2024}. 

\begin{figure*}[htb]
\centering
\includegraphics[width=0.9\linewidth]{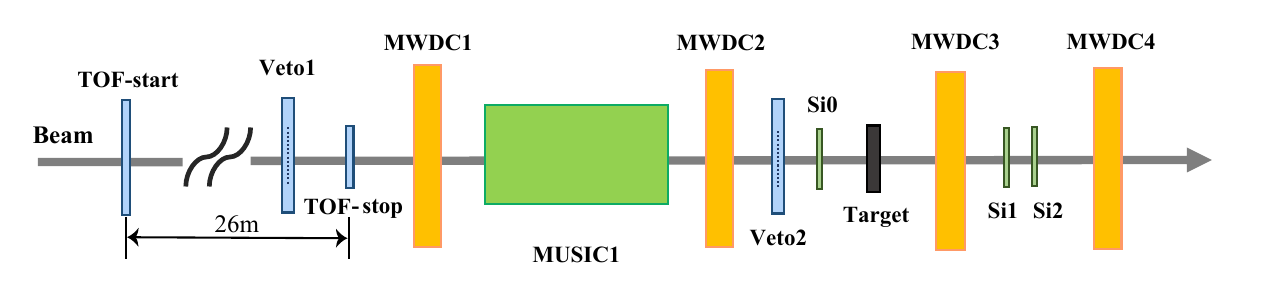}
\caption{(Color online) Layout of the scheme of detector setup for the CCCS measurements in the ETF (not to scale).} \label{f1}
\end{figure*}

Recent efforts have been made to extract information on the point-proton radii of light unstable nuclei from precise CCCS data~\cite{Terashima2014,Tran16,ZhangJC2024,ZhaoJW2024,Imran2024}, to probe neutron skin thickness~\cite{Yamaguchi2011,LiXF2016,Kanungo2016,Bagchi2019,Ozawa2014,Estrad2014,Kaur2022} and halo structures~\cite{Kanungo2016,Bagchi2019}. These investigations were carried out primarily for light nuclei on carbon (C) and hydrogen (H) targets at energies below 1 GeV/nucleon. Glauber-type models~\cite{MengJ2002, Bhagwat2004} were used to calculate the CCCS based on the direct projectile proton removal process, governed by the strong nuclear interaction between nucleons. In this framework, only the protons of the projectile interacting with the target nucleus contribute to the CCCS, whereas the neutrons of the projectile are treated as spectators. However, detailed investigations have shown that theoretical results from the Glauber model calculation systematically underestimate the experimental data~\cite{Yamaguchi2010,Yamaguchi2011,Tanaka2022,ZhangJC2024,ZhaoJW2023,WangCJ2023}. Charged particle evaporation (CPE) after neutron removal has recently been found to be crucial to resolving this discrepancy between Glauber model predictions and experimental data on hydrogen and carbon~\cite{Estrad2014,Bagchi2019,Tanaka2022,ZhaoJW2023,ZhangJC2024,Tanaka2024}. Moreover, this contribution is not identical but isospin-dependent with the maximum at isospin symmetric nuclei~\cite{ZhaoJW2023,ZhangJC2024}, e.g., $^{10,11}$B~\cite{Estrad2014} and $^{14,15}$N~\cite{Bagchi2019}. Nevertheless, since all these studies mentioned above have focused on light targets, the contribution of electromagnetic (EM) interaction to the CCCS on heavy targets below 1~GeV/nucleon remains uncertain.

Early fragmentation cross section studies of $^{12}$C and $^{16}$O at 1.05 and 2.1~GeV/nucleon on various target materials demonstrated significant EM contribution for heavy targets~\cite{Heckman1976}. Subsequent experiments at above 1~GeV/nucleon revealed that the EM contribution increases with increasing reaction energies~\cite{Hirzebruch1992,Hill1991,Schei02,Brechtmann1988-1}, projectiles~\cite{Hill1988,Hill1991} and targets~\cite{Hirzebruch1992,Schei02,Schei04,Westfall1979,Hirzebruch1995,Olson1981,Brechtmann1988-1} of higher Z. For example, the contribution of electromagnetic dissociation (EMD)  accounts for 6\% to approximately 40\% of the CCCS for $^{16}$O on a lead (Pb) target as the energy increases from 2.3 to 200 GeV/nucleon \cite{Hirzebruch1992}. The enhancement of the inclusive nuclear-charge pickup cross sections for 158 GeV/nucleon $^{208}$Pb on targets higher Z than Aluminum, compared to the cross sections at 10.6 GeV/nucleon, has been attributed to EMD processes involving pion production~\cite{Schei02}. 
Nowadays, EMD at a few hundred MeV/nucleon is considered as an established tool for determining the EM response of weakly bound nuclei~\cite{Bertulani2023, Cook2020}. However, systematic data and investigations of the EMD effect on CCCSs at energies below 1 GeV/nucleon remain scarce.

In this work, we report the CCCS measurements of $^{18}$O on carbon and lead targets at energies around 370~MeV/nucleon. After introducing the experiment and data analysis in Section~\ref{expana}, we present the experimental results and show the evaluation of contributions from the nucleon-nucleon (NN) and EM interactions to CCCSs in Section~\ref{resdis}. Finally, a summary is given in Section~\ref{sum}.

\section{Experiment and data analysis}
\label{expana}

The experiment was performed at the second Radioactive Ion Beam Line at Lanzhou (RIBLL2) at the Heavy Ion Research Facility (HIRFL)~\cite{zhou2016heavy}, China. A primary beam of $^{18}{\rm O}$ at 400~MeV/nucleon from the CSRm was transported by the first half of the RIBLL2 to the reaction target installed at the External Target Facility (ETF)~\cite{Sun_2018}. Figure~\ref{f1} shows the schematic diagram of the experimental setup installed at ETF. The time-of-flight (TOF) of the incident particle was measured with a plastic scintillation counter (TOF-start) at the first focal plane of RIBLL2 and anther plastic scintillation counter (TOF-stop) at ETF. The TOF resolution achieved was better than 100 ps ($\sigma$)~\cite{lin2017plastic}. Upstream of the reaction target, a Frisch grid type multiple sampling ionization chamber (MUSIC1) was sandwiched between two multi-wire drift chambers (MWDC1 and MWDC2) to measure the energy loss ($\Delta E$) of incident particles~\cite{zhang2015multiple, ZhaoJW2019}. The active area of MUSIC1 is 85~$\times$~85~mm$^{2}$. A silicon counter (Si0) with a thickness of 300~$\mu$m was also installed as a redundant $\Delta E$ detector. Additionally, two 100~$\times$~100~mm$^{2}$ plastic scintillation counters (Veto1 and Veto2), each with a 30~$\times$~30~mm$^{2}$ aperture in the center, were positioned upstream of the reaction target. The two active collimators were used to limit the size of the beam spot on the target in addition to the tracking information from the MWDCs.

Downstream of the reaction target, two silicon counters (Si1 and Si2), each with a resolution of approximately 0.2 ($\sigma$) and the same thickness as Si0, were positioned between two corresponding multi-wire drift chambers (MWDC3 and MWDC4) to measure the $\Delta E$ of outgoing particles for their $Z$ identification. The particle tracking before and after the reaction target was obtained by the MWDCs with a sensitive area of 130~$\times$~130~mm$^{2}$. The position resolution is about 120~$\mu$m ($\sigma$)~\cite{ZhaoJW2018}, and the detection efficiency is approximately 95\%. A 2.767 (2)~g/cm$^{2}$ natural carbon ($^{\text{nat}}$C) target and a 3.797 (3)~g/cm$^{2}$ natural lead ($^{\text{nat}}$Pb) target were used. The energies in the middle of the $^{\text{nat}}$C and the $^{\text{nat}}$Pb targets are 372 (1)~MeV/nucleon and 378 (1)~MeV/nucleon, respectively. 

CCCSs were measured with the transmission method as
\begin{equation}\label{eq1}
\sigma_\text{cc}= \frac{1}{t}\ln(\frac{\gamma ^{0}}{\gamma}) \;,
\end{equation}
where $t$ is the number of target nuclei per unit area, $\gamma=N_\text{out}/N_\text{in}$ represents the transmission probability under the reaction target condition. $N_\text{in}$ is the number of incidents $^{18}{\rm O}$ and $N_\text{out}$ is the number of particles without losing any protons after reactions. To eliminate the effect of reactions in materials other than the reaction target, measurement was also performed without the reaction target to get $\gamma^{0}=N_\text{out}^{0}/N_\text{in}^{0}$. The energy differences arising from the presence of a target versus an empty target have a negligible impact on the deduced cross sections. 

\begin{figure}[htbp]
\centering
\includegraphics[width=0.95\linewidth]{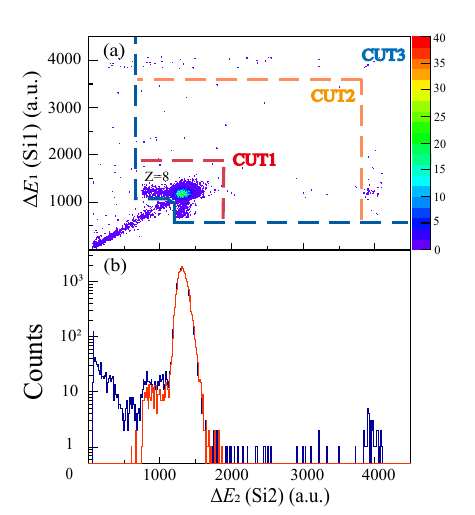}
\caption{ (color online) (a) Charge identification of outgoing particles for Pb target with $^{18}$O as incident particles. CUT1 defines the outgoing residual nuclei with a proton number of $Z$-unchanged and Z+1. Z+1 particles come from charge-pickup or charge-exchange reactions~\cite{Tanihata2016}. Two additional selections (CUT2 and CUT3) are considered for systematic uncertainties estimation. (b) Projection of the x-axis from panel (a). This corresponds to the energy deposition of the incident particles $^{18}$O in Si2 after the Pb target. The blue line in (b) represents the energy deposition of all outgoing particles in Si2, (b) represents the energy deposition of all outgoing particles in Si2, while the red line represents the case with selecting $Z=$ 8 and 9 particles by gating CUT1.}
\label{f3}
\end{figure}

The key to determine the CCCS is to count incident particles before the reaction target and the $Z$-unchanged outgoing particles after the reaction target. The TOF-magnetic rigidity-$\Delta E$ method was used to identify and count incident $^{18}$O particles event-by-event before the reaction target. Tracking information from MWDC1 and MWDC2 is used to define the beam diameter of $\phi$33 mm on reaction targets ensuring the beam size is smaller than the target area of $\phi$50 mm. This configuration guarantees that Si1 and Si2 will fully accept outgoing particles with a charge number of $Z=$ 8. As the example of the Pb case shown in Fig.~\ref{f3}, to resolve the channeling effect, the $Z$-unchanged particles were identified using the two-dimensional $\Delta E_{\text{Si1}} - \Delta E_{\text{Si2}}$ spectrum. With Veto1 and Veto2 applying additional selection of the incident particles, we found that the experimental data obtained with such additional selection were lower by 0.1\% on $^{{\text{nat}}}$C and 0.7\% on $^{{\text{nat}}}$Pb than without considering the information from Veto1 and Veto2. The difference is much smaller than the statistic uncertainty, thus we use the cross sections without veto information hereafter for discussions. Further details on the data analysis procedures can be found in Ref.~\cite{WangCJ2023}.

\section{Results and discussions}
\label{resdis}

CCCSs of $^{18}$O on $^{\text{nat}}$C and $^{\text{nat}}$Pb measured in this work are summarized in Table~\ref{tab2}. The CCCS on the $^{{\text{nat}}}$Pb target is approximately four times larger than that on the $^{{\text{nat}}}$C target. Our CCCS result on C agrees in error bars with the previous data at similar reaction energies~\cite{Yamaguchi2011,ZhaoJW2023}. The direct proton removal process, along with the CPE after neutron removal and the EMD, contributes to CCCS. We estimate these three components in detail as follows.

\subsection{Direct proton removal and charged particle evaporation}

\begin{table*}[t]
    \centering
    \caption{Contribution of $\sigma_\text{cc}^\text{direct}$, $\sigma_\text{cc}^\text{evap}$, and $\sigma_\text{cc}^\text{EM}$, to the CCCSs ($\sigma_\text{cc}^\text{cal}$) in comparison with the experimental data $\sigma_\text{cc}^\text{exp}$ for $^{18}$O beam on $^\text{nat}$C and $^\text{nat}$Pb targets. The $\sigma_\text{cc}^\text{EM}$ are determined by $b_\text{min}$ from Eq.(\ref{eq7}), with uncertainties derived from the differences between Equations Eq.(\ref{eq7}) and Eq.(\ref{eq8}).
    }
    \label{tab2}
    \begin{ruledtabular}
    \begin{tabular}{lccccccr}
    \text{Projectile + Target} & \text{Energy (MeV/nucleon)} & \text{$\sigma_\text{cc}^\text{direct}$(mb)} & \text{$\sigma_\text{cc}^\text{evap}$(mb)} & \text{$\sigma_\text{cc}^\text{EM}$(mb)} & \text{$\sigma_\text{cc}^\text{cal}$(mb)} & \text{$\sigma_\text{cc}^\text{exp}$(mb)} \\
    \hline
    $^{18}$O+$^\text{nat}$C & 372 (1) &  764 & 106 & 0.2 (19) &  870.2 (19) &  863 (21) \\ 
    $^{18}$O+$^\text{nat}$Pb & 378 (1) &  3169 & 169 & 9.2 (49) & 3347.2 (49) & 3386 (137) \\
    \end{tabular}
    \end{ruledtabular}
\end{table*}

The direct proton removal cross sections, $\sigma_{\text{cc}}^{\text{direct}}$, is calculated with the zero-range optical-limit approximation Glauber model (ZRGM)~\cite{MengJ2002,Bhagwat2004}. In the case of $^{\text{nat}}$C target, $^{12}$C is used in the calculation instead of the $^{\text{nat}}$C as it is predominantly (approximately 99\%) composed of $^{12}$C. Proton and neutron density distributions based on the Harmonic-oscillator (HO) model are employed for both $^{12}$C and $^{18}$O. 
\begin{equation}\label{eq20}
\rho(r) = \rho_{0}\left[1+\frac{C-2}{3}\left(\frac{r}{\omega}\right)^{2}\right]exp\left[-\left(\frac{r}{\omega}\right)^{2}\right],
\end{equation}
where $\rho_{0}$ denotes the central density, $C$ represents the number of protons or neutrons, and $\omega$ is the radius parameter. The relevant density distributions are determined by reproducing the experimental charge radii~\cite{Angeli2013} and the interaction cross sections on carbon at around 1000 MeV/nucleon~\cite{Ozawa2001}. Detailed parameters used in the nucleon density distributions of $^{12}$C and $^{18}$O are listed in Table~\ref{tab1}.

\begin{table}[htbp]
    \centering
    \caption{Nuclear proton and neutron density parameters and the corresponding root-mean-square radius ($R_{p}$ and $R_{n}$) for $^{18}$O and $^{12}$C.}
    \label{tab1}
    \begin{ruledtabular}
    \begin{tabular}{lccccccr}
        \textrm{Nucleus} & \textrm{$\omega_{p}$} & \textrm{$\omega_{n}$} & \textrm{$\rho_{0p}$} & \textrm{$\rho_{0n}$} & \textrm{$R_{p}$}  & \textrm{$R_{n}$} \\
        & \textrm{(fm)} & \textrm{(fm)} & \textrm{(fm$^{-3}$)} & \textrm{(fm$^{-3}$)} & \textrm{(fm)} & \textrm{(fm)} \\
        \hline
        $^{18}$O & 1.768 & 1.716 & 0.065 & 0.071 & 2.652 &  2.603 \\ 
        $^{12}$C & 1.582 & 1.569 & 0.091 & 0.093 & 2.328 &  2.309 \\ 
    \end{tabular}
    \end{ruledtabular}
\end{table}
 
In the case of $^{\text{nat}}$Pb target, it consists of $^{204}$Pb, $^{206}$Pb, $^{207}$Pb, and $^{208}$Pb. The abundance of $^{204}$Pb is negligible (1.4\%), while the abundance ratio of $^{206}$Pb, $^{207}$Pb, and $^{208}$Pb is about 1:1:2. The proton and neutron density distributions for $^{204}$Pb, $^{206}$Pb, $^{207}$Pb and $^{208}$Pb are obtained using the Skyrme-Hartree-Fock models with the SkM* force~\cite{Bartel1982}. This force successfully explains both the electron scattering data~\cite{WA2003} and nucleon-nucleus elastic scattering data for $^{208}$Pb~\cite{SK2002}. Variations in the Skyrme forces result in a negligible impact on the cross sections. The final cross section for the targets $^{\text{nat}}$Pb is determined by taking the weighted average of the calculated cross sections for $^{204}$Pb, $^{206}$Pb, $^{207}$Pb, and $^{208}$Pb in ZRGM according to their respective abundances. Replacing $^{\text{nat}}$Pb with $^\text{208}$Pb will lead to a change of the CCCS within about 10 mb.  
We also performed calculations with the finite-range optical-limit approximation Glauber model (FRGM)~\cite{Abu-Ibrahim2008}. Comparing with the results from the ZRGM, the FRGM predicts about 1.3\% and 0.8\% larger cross sections of $^{18}$O on the $^{\text{nat}}$C and  $^{\text{nat}}$Pb targets, respectively. We adopt the ZRGM results, $\sigma_\text{cc}^\text{direct}$ as summarized in Table~\ref{tab2} in the following discussion.

The CPE cross sections, $\sigma _\text{cc}^\text{evap}$, are computed following the method described in Ref.~\cite{ZhaoJW2023}. 
The only input parameter in this calculation is the maximum excitation energy ($E_\text{max}$) of single-neutron-removed pre-fragments. For $^{18}$O, $E_\text{max}$ is set at 42 MeV, which takes into account both the energy of a single particle hole relative to the Fermi surface within the framework of the Fermi gas model~\cite{Tanaka2022,GS1991,AM2010} and the experimental CCCS value reported in Ref.~\cite{ZhaoJW2023}. Adjusting $E_\text{max}$ by 30\% has a negligible effect on $\sigma _\text{cc}^\text{evap}$, causing CCCSs to change by less than 3\%. The cross sections from the CPE after neutron removal, $\sigma_\text{cc}^\text{evap}$, are summarized in Table~\ref{tab2}. 

\subsection{Electromagnetic dissociation}

When a swift projectile nucleus passes by the heavy target nuclei, it experiences the virtual-photon field generated by the target nuclei. Such virtual photons can excite the projectile nucleus~\cite{Norbury2007}, thus contributing to the CCCS.  
Depending on its energy, a virtual photon can be absorbed by a nucleus through different mechanisms. For virtual photons with energies of $E_{\gamma}\leq$ 40~MeV, the excitation of a nucleus as a whole in the form of a giant dipole resonance (GDR) is the most probable photoabsorption process. Virtual photons within the energy interval of 40 $\leq E_{\gamma}\leq$ 140~MeV allow for photon absorption by the quasi-deuterons ($i.e.$, correlated proton-neutron pairs) in a nucleus~\cite{IAEA2020}. The pion photoproduction occurs when $E_{\gamma}$ exceeds 140~MeV~\cite{Pshenichnov2011}.

The excited projectile nucleus is likely to break up by emitting light-charged particles, thus contributing to the CCCS.
This EM component, $\sigma _\text{cc}^\text{EM}$, can be expressed as:
\begin{equation}\label{eq5}
\sigma_\text{cc}^\text{EM} = \int_{S_{\alpha}}^{E_{\gamma}^\text{max}} N( E_{\gamma} ) \sigma_{\gamma}^\text{charge} (E_{\gamma})  dE_{\gamma} \; ,
\end{equation}
where $N(E_{\gamma})$ and $\sigma_{\gamma}^\text{charge}(E_{\gamma})$ represent the total number of virtual photons with the energy of $E_{\gamma}$ and the photonuclear reaction cross section, respectively. 
The $\alpha$ is the most easily emitted charged particle from $^{18}$O, with an emission threshold $S_\alpha$ = 6.228 MeV. The maximum energies of the virtual photons are often estimated by $E_{\gamma}^\text{max}=\gamma \beta\hbar c/b_\text{min}$~\cite{Bertulani1988}, where $\beta = v / c $, $v$ is the velocity of the target, $c$ is the speed of light, and $\gamma = 1 / \sqrt { 1 - \beta ^ { 2 } } $.
$b_\text{min}$ is the minimum impact parameter. 

We tried two ways to estimate $b_\text{min}$.
In the first method as in Ref.~\cite{Olson1981}, $b_\text{min}$ can be calculated as
\begin{equation}\label{eq7}
b_\text{min} = R_{0.1}(P) + R_{0.1}(T) -d \;,
\end{equation} 
where $R_{0.1}(P)$ and $R_{0.1}(T)$ represent the charge radius of the projectile and target nucleus at 10\% of the central density, respectively. $d$ is fixed to -1.5~fm by fitting experimental values~\cite{Olson1981}. The second estimation of 
$b_\text{min}$ is following Ref.~\cite{Benesh1989}, and is calculated by 
\begin{equation}\label{eq8}
b_\text{min} = r_{0}[A^{1/3}+B^{1/3}-X(A^{-1/3}+B^{-1/3})]\; ,
\end{equation} 
with $r_{0}$ = 1.34 fm, $X$ = 0.75. $A$ and $B$ represent the mass numbers of the projectile and the target, respectively. $b_\text{min}$ derived from Eq.(\ref{eq8}) can vary by up to 19\% for Pb and a factor of 2 for C compared to that from Eq.(\ref{eq7}). This results in a change of about 53\% and a factor of 10 in $\sigma_\text{cc}^\text{EM}$ for Pb and C, respectively. Nevertheless, the $\sigma_\text{cc}^\text{EM}$/$\sigma_\text{cc}^\text{exp}$ ratio shows only a small variation of 0.15\% for Pb and 0.22\% for C. 

\subsubsection{Virtual photons spectra }

Overall, the dominant transitions in the EM process are E1 and E2. The multipolar virtual photon spectra are calculated using the equivalent photon method~\cite{Bertulani1988} as:
\begin{equation}
\label{eq4}
\begin{aligned}
N_\text{E1} (  E_{\gamma} ) = \frac { 2 Z  ^{ 2 } \alpha  } { \pi E_{\gamma} \beta^ { 2 }}  [ \xi K _ { 0 }(\xi) K _ { 1 }(\xi) ] \\
- \frac { 1  } { 2  }\xi ^ { 2 } \beta ^ { 2 } ( K _ { 1 }(\xi) ^ { 2 } - K _ { 0 }(\xi) ^ { 2 } ) \;,
\end{aligned}
\end{equation}

\begin{equation}
\label{eq4}
\begin{aligned}
N_\text{E2} ( E_{\gamma} ) = \frac { 2\alpha Z  ^ { 2 }} { \pi E_{\gamma} \beta ^ { 4 }}     [2(1-\beta ^ { 2 })( K _ { 1 }(\xi) ^ { 2 }+\xi(2-\beta ^ { 2 })^{2}
\\K _ { 0 }(\xi)K _ { 1 }(\xi)-\frac{\xi^{2}\beta ^ { 4 }}{2}( K _ { 1 }(\xi) ^ { 2 } - K _ { 0 }(\xi) ^ { 2 } )] \;,
\end{aligned}
\end{equation}
where $Z$ is the number of protons in the target nucleus, $\alpha$ is the structural fine constant, $K_{0}$ and $K_{1}$ are the zeroth and first terms of the first kind of Bessel function, respectively. $\xi$ is defined as
\begin{equation}
\label{eq6}
\xi \equiv  \frac{ E_{\gamma} b_\text{min}}{\gamma \beta \hbar c } .
\end{equation} 

\begin{figure}[htbp]
\centering
\includegraphics[width=1\linewidth]{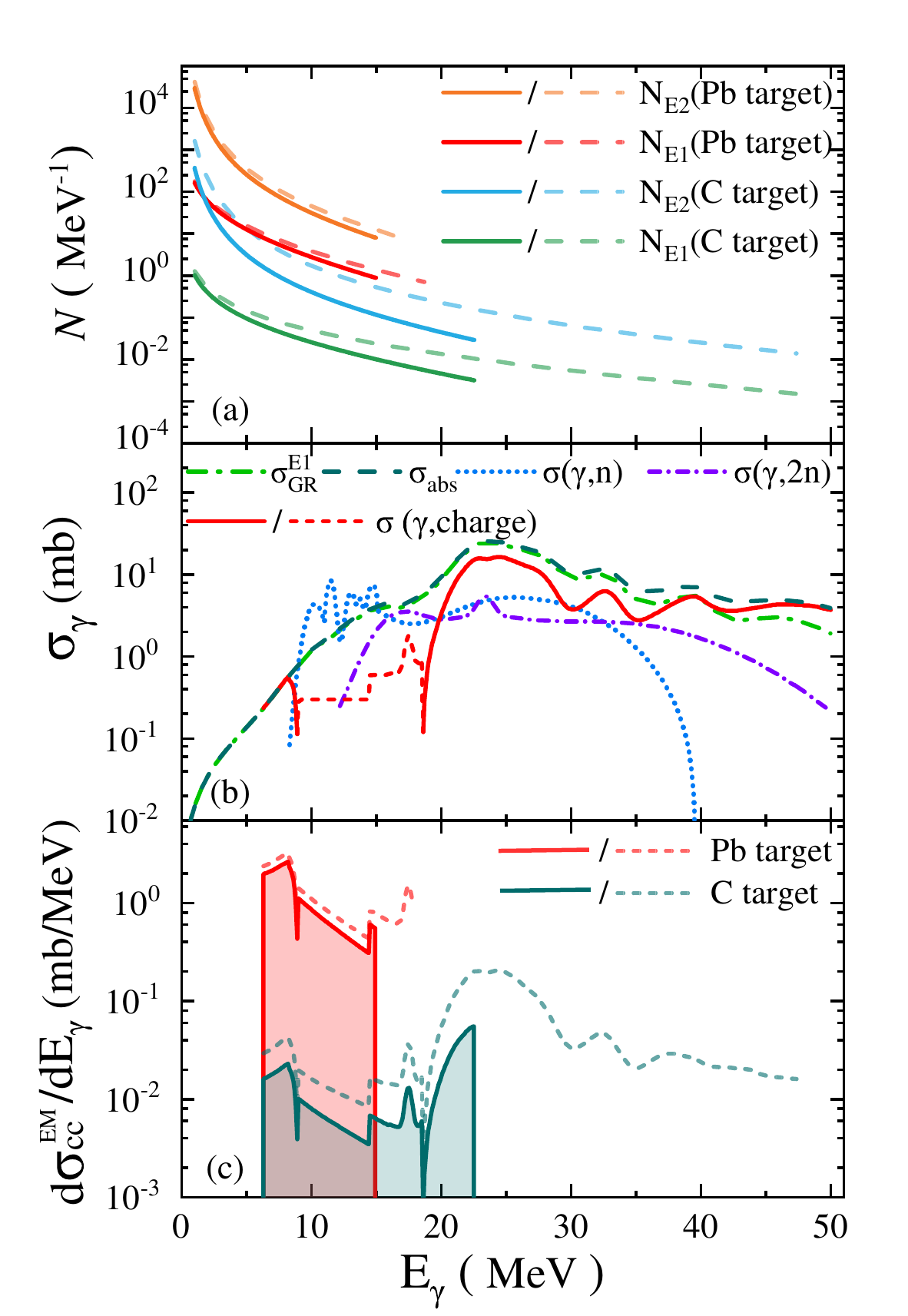}
\caption{
(a) E1 and E2 virtual-photon spectrum, 
$N(E_{\gamma})$, for two different impact parameters on C and Pb targets. 
Shown are the cases calculated using $b_\text{min}$ from Eq.~(\ref{eq7}) (solid line) and Eq.~(\ref{eq8}) (dashed line). (b) Photon-induced cross sections of $^{18}$O.
(c) Differential EM cross sections calculated with two different $b_\text{min}$. The shadowed areas represent the resulting $\sigma_\text{cc}^\text{EM}$ for C and Pb targets. For details, refer to the text.}
\label{f5}
\end{figure}

As shown in Fig.~\ref{f5} (a),  $N_\text{E1}(E_{\gamma})$ and $N_\text{E2}(E_{\gamma})$ calculated with the two $b_\text{min}$ parameteres, are monotonically decreasing with $E_{\gamma}$, but $N_\text{E2}(E_{\gamma})$ is two orders of magnitude greater than $N_\text{E1}(E_{\gamma})$ at the same $E_{\gamma}$ for the same target. Moreover, both $N_\text{E2}(E_{\gamma})$ and $N_\text{E1}(E_{\gamma})$ of the Pb target are larger than that of the C target by two orders of magnitude.

\subsubsection{Photonuclear reaction cross sections}

The photoabsorption cross section $\sigma_\text{abs}$ can be described as 
\begin{equation}\label{eq11}
\sigma_\text{abs} = \sigma_\text{GR}^\text{E1} + \sigma_\text{GR}^\text{E2} +\sigma_\text{QD} ,
\end{equation}
where $\sigma_\text{GR}^\text{E1}$ and $\sigma_\text{GR}^\text{E2}$ represent the giant resonance (GR) cross sections corresponding to E1 and E2 transitions, respectively.
$\sigma_\text{QD}$ denotes the quasi-deuteron (QD) cross section. $\sigma_\text{GR}$ can be derived from the microscopic quasiparticle random-phase approximation (QRPA) utilizing the SV-M56-O Skyrme interaction. This method has been proven to offer a better description of $\sigma_\text{abs}$ for $^{16}$O~\cite{Lyutorovich2012}. $\sigma_\text{QD}$ is computed using the Chadwick's model~\cite{Chadwick1991}.
We plot the cross sections $\sigma_\text{GR}^\text{E1}$ and $\sigma_\text{abs}$ in Fig.~\ref{f5} (b). For the incident energies $E_{\gamma}$ below 50 MeV, it is evident that $\sigma_\text{GR}^\text{E1}$ represents the major fraction of $\sigma_\text{abs}$.

In the photo-neutron cross section of $^{18}$O, $\sigma(\gamma, n)$ and $\sigma(\gamma, 2n)$ are the dominant contributions. Therefore, $\sigma_{\gamma}^{\text{charge}}$ can be calculated as
\begin{equation}\label{eq10}
\sigma_{\gamma}^\text{charge} \approx \sigma_\text{abs} - \sigma (\gamma , n) - \sigma (\gamma , 2n) ,
\end{equation}
where $\sigma (\gamma, n)$ and $\sigma (\gamma, 2n)$ are obtained from the existing experimental data~\cite{Woodworth1979}.
One should note that the experimental determination of $\sigma(\gamma, n)$ incorporates a contribution from $\sigma(\gamma, np)$~\cite{Woodworth1979,McNeill1991}, which necessitates the exclusion of this component.
The pure $\sigma(\gamma, n)$ and $\sigma(\gamma, 2n)$ cross sections are plotted in Fig.~\ref{f5} (b).

Within the energy range of 9-18.5 MeV, the calculated $\sigma_{\text{abs}}$ using Eq.~(\ref{eq5}) is found to be smaller than the experimental $\sigma(\gamma, n)$. Therefore, instead of using $\sigma_{\text{abs}}$, we utilize the available experimental data for $\sigma(\gamma, \alpha)$ from 9 to 18.5 MeV~\cite{Bangert1982}, $\sigma(\gamma, n\alpha)$ from 14.5 to 18.5 MeV, and $\sigma(\gamma, p)$ from 16 to 18.5 MeV~\cite{Woodworth1979}. For $\sigma(\gamma, \alpha)$ and $\sigma(\gamma, n\alpha)$, an upper limit of 0.3 mb is applied. The adopted $\sigma_{\gamma}^\text{charge}$ in 9-18.5~MeV is also shown by the dashed line in Fig.~\ref{f5} (b).

\subsubsection{EMD contributions to CCCS}

The weighted virtual photon spectra can be safely given by $N(E_{\gamma}) = M \cdot N_\text{E1}(E_{\gamma}) + (1-M) \cdot N_\text{E2}(E_{\gamma})$, where $M$ is based on $\sigma_\text{abs}$. Previous studies have shown that the $E1$ term accounts for nearly the entire total photoelectric cross sections within the QD region for photon energies from 20 MeV to 140 MeV~\cite{JFM1949-1,JFM1949-2,LIS1950}. Therefore, we have $M = (\sigma_\text{GR}^\text{E1} + \sigma_\text{QD}) / \sigma_\text{abs}$. 

In this way, we have both $N(E_{\gamma})$ and $\sigma_{\gamma}^\text{charge}$ in Eq.~(\ref{eq5}). Multiplying these two terms and integrating over the range $S_{\alpha}$ to $E_{\gamma}^\text{max}$ will yield  $\sigma_{\text{cc}}^{\text{EM}}$. As shown in Fig.~\ref{f5} (c), the red and teal shaded regions represent the final $\sigma_{\text{cc}}^{\text{EM}}$ estimations of $^{18}$O on Pb and C targets, 
respectively. These results are obtained with the $b_\text{min}$ determined with Eq.~(\ref{eq7}) and listed in the Table~\ref{tab2}.  For comparison, we have performed another calculation using $b_\text{min}$ from Eq.~(\ref{eq8}), as shown by the dashed lines in Fig.~\ref{f5} (a) and (c).

Estimations of $\sigma_\text{cc}^\text{direct}$, $\sigma_\text{cc}^\text{evap}$, and $\sigma_\text{cc}^\text{EM}$, along with their sum $\sigma_\text{cc}^\text{cal}$, are presented in Table~\ref{tab2}. Both $\sigma_\text{cc}^\text{exp}$ of $^{18}$O on the C and the Pb targets measured in this work can be reproduced well by considering the contribution from the direct proton removal process, the CPE after neutron removal and the EMD. $\sigma_\text{cc}^\text{EM}$ and $\sigma_\text{cc}^\text{evap}$ on Pb account for only 5.3\% of the experimental CCCS but for 12.3\% on C target.

\subsubsection{Prediction for reaction systems of higher Z} 

Precise CCCS measurements were performed mainly at energies around 300 and 900~MeV/nucleon. Therefore, we have applied the current method to investigate $\sigma_{\text{cc}}^{\text{EM}}$ of $^{18}$O, $^{59}$Co, $^{112}$Sn, $^{154}$Sm, and $^{197}$Au projectiles striking on the C, silver (Ag) and Pb targets at 300~MeV/nucleon and 900~MeV/nucleon. 

For the $\sigma_{\text{cc}}^{\text{EM}}$ estimation, the $\sigma_\text{abs}$ values of $^{59}$Co, $^{154}$Sm, and $^{197}$Au are taken from the measurements in Refs.~\cite{Wyckoff1965,Gurevich1981}. In the case of $^{112}$Sn, $\sigma_\text{abs}$ is calculated by the quasiparticle-vibration coupling (QPVC) model due to the lack of experimental results. This model has been proved to give a good description to the $\sigma_\text{abs}$ of Sn isotopes~\cite{LiZZ2024}. Experimental $\sigma(\gamma, n)$ and $\sigma(\gamma, 2n)$ of $^{59}$Co, $^{112}$Sn, $^{154}$Sm, and $^{197}$Au are taken from  Refs.~\cite{Alvarez1979,Veyssiere1970,Varlamov2010}.
For all projectile-target systems, $b_\text{min}$ are calculated with Eq.~(\ref{eq8}). As the E1 transition is dominant, we only consider $N_\text{E1}(E_{\gamma})$ produced by the target here in the $\sigma_{\text{cc}}^{\text{EM}}$ calculation. 

To calculate $\sigma_{\text{cc}}^{\text{direct}}$ and $\sigma_{\text{cc}}^{\text{evap}}$, the proton and neutron density distributions from the two-parameter Fermi model have been adopted for $^{59}$Co, $^{112}$Sn, $^{154}$Sm, $^{197}$Au and $^{107}$Ag. The relevant $E_\text{max}$ values were assumed to be the maximum Fermi energies. Varying $E_\text{max}$ values by 30\% leads to a change of $\sigma_{\text{cc}}^{\text{evap}}$ by less than 2.6\%, which has a negligible effect on the  $\sigma_{\text{cc}}^{\text{EM}}/\sigma_{\text{cc}}^{\text{cal}}$ ratios. The experimental data for $^{18}$O on C target at around 900 MeV/nucleon in Ref.~\cite{Kaur2022} is 8\% lower than that in Ref.~\cite{Chulkov2000} and our predicted $\sigma_{\text{cc}}^\text{cal}$ agree well with the Ref.~\cite{Chulkov2000}.

\begin{figure}[h]
\centering
\includegraphics[width=1\linewidth]{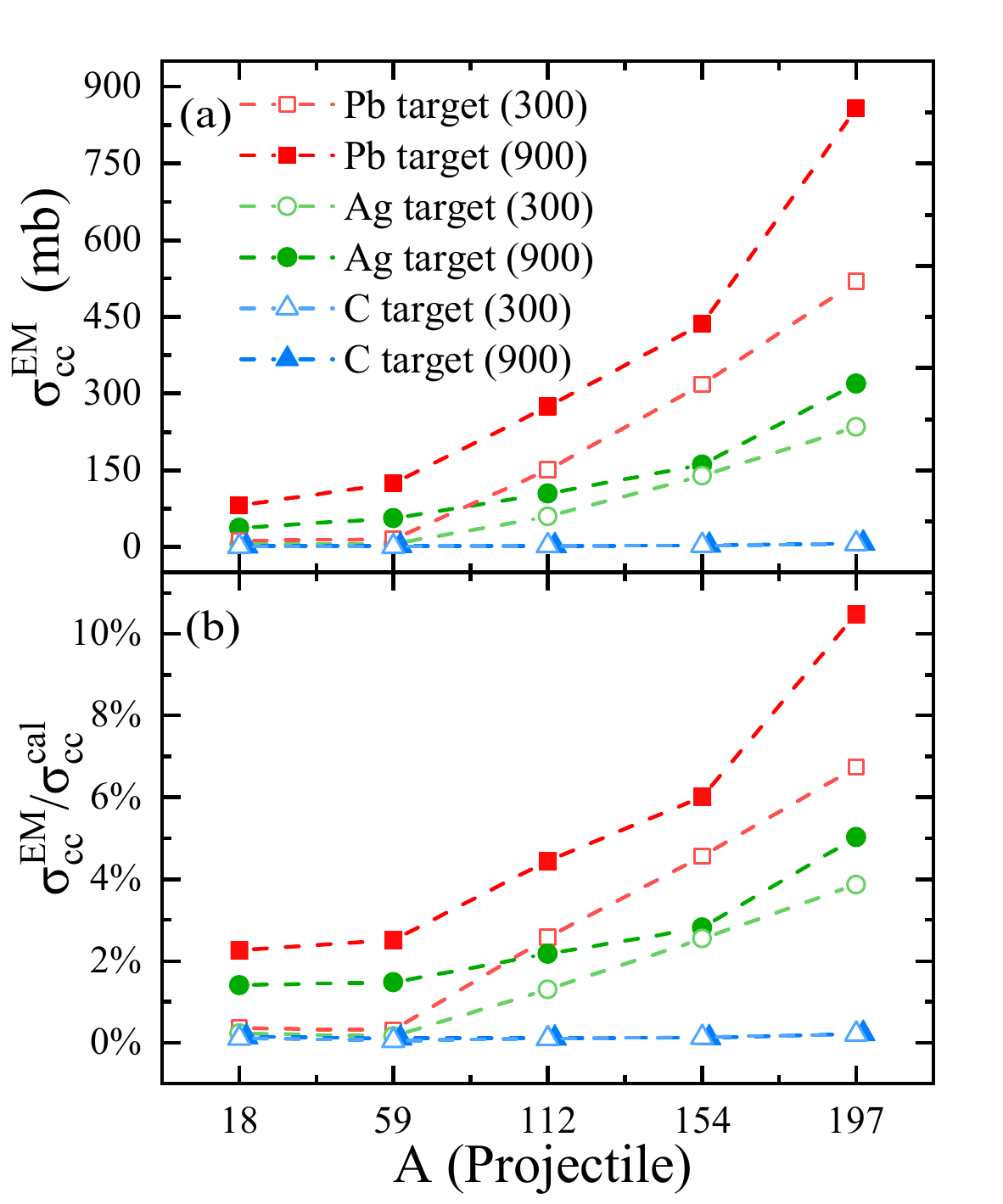}
\caption{$\sigma_{\text{cc}}^{\text{EM}}$ and $\sigma_{\text{cc}}^{\text{EM}}$/$\sigma_{\text{cc}}^\text{cal}$ ratios calculated for $^{18}$O, $^{59}$Co, $^{112}$Sn, $^{154}$Sm and $^{197}$Au projectiles on C, Ag, and Pb targets at 300 and 900~MeV/nucleon. To distinguish $\sigma_{\text{cc}}^{\text{EM}}$ and $\sigma_{\text{cc}}^{\text{EM}}$/$\sigma_{\text{cc}}^\text{cal}$ ratios of different projectiles at two energies on C, the solid triangle is shifted slightly to the right. Lines are only for guiding eyes.
}
\label{f7}
\end{figure}

$\sigma_{\text{cc}}^{\text{EM}}$ and $\sigma_{\text{cc}}^{\text{EM}}/\sigma_{\text{cc}}^{\text{cal}}$ ratios are dependent on the incident energy and projectile-target combinations, as shown in Fig.~\ref{f7}. The EM contribution to the CCCS is negligible for all projectiles when interacting with the C target at incident energies of 300 and 900~MeV/nucleon. 
In cases of projectiles impinging on Ag and Pb targets, the ratios exhibit an increasing trend towards the heavy projectiles. The reason for such a trend is that $\sigma_\text{abs}$ at lower $E_{\gamma}$ is sensitive with projectile mass number~\cite{Skobel1967}, enhancing $\sigma_\text{cc}^\text{EM}$ from $^{59}$Co to $^{197}$Au.
In general, the ratios of all projectiles at 900 MeV/nucleon are systematically larger than those at 300 MeV/nucleon due to the higher total virtual photon production yield and increasing photonuclear reactions with higher energy~\cite{Bertulani1988}. The EM affects weakly to the CCCS of projectiles lighter than $^{59}$Co on both Ag and Pb targets at 300~MeV/nucleon. With increasing either the incident energy or the projectile mass, the ratios can exceed over 1\%. For instance, the ratios are 6.7\% and 10.5\% for $^{197}$Au on Pb at 300 and 900 MeV/nucleon, respectively. 

\section{Summary}
\label{sum}

We have measured CCCSs of $^{18}$O on C and Pb targets with an uncertainty of less than 4\% at around 370 MeV/nucleon. The experimental CCCSs were well reproduced by considering the contributions of the direct proton removal process, the CPE after neutron removal and the EM interaction. Our results indicate that the CPE following the projectile neutron removal is crucial for understanding the experimental data on both the light C and the heavy Pb targets. Specifically, the CPE contribution to the CCCSs of $^{18}$O on C and Pb targets is 12.3\% and 5\%, respectively. The influence of EM interaction on the CCCS in the current energy regime is less than 1\%. Therefore, the estimation of CCCSs on the Pb target has a better tolerance for the uncertainty of calculating $\sigma_{\text{cc}}^\text{evap}$ and $\sigma_{\text{cc}}^\text{EM}$ than the case on the C target. 

Moreover, we conducted further investigations of $^{18}$O, $^{59}$Co, $^{112}$Sn, $^{154}$Sm and $^{197}$Au projectiles on C, Ag, and Pb targets at 300 and 900~MeV/nucleon. It turns out that the contribution of EM to the CCCS on Ag and Pb increases significantly with the projectile mass and the incident energy. The EM contribution to the CCCS of $^{197}$Au on Pb at 300 and 900~MeV/nucleon reaches 6.7\% and 10.5\%, respectively. In contrast, the EM contribution is negligible for the CCCSs of all projectiles on the C target.

\section*{Acknowledgments}
We thank the HIRFL-CSR accelerator team for their efforts to provide a stable beam condition during the experiment. We thank Dr. Arjan Koning and Dr. Yi Xu for the discussions regarding photonuclear cross sections. This work was supported partially by the National Natural Science Foundation of China (Nos. 12325506, 11961141004, 11922501), and the ``111 Center" (Grant No. B20065).


\end{document}